\documentclass[aps,floats,twocolumn,showpacs,showkeys]{revtex4}
\usepackage{graphicx}
\usepackage{epsfig}
\input{psfig.sty}

\newcommand{\bastar}{\begin{eqnarray*}}
\newcommand{\eastar}{\end{eqnarray*}}
\newskip\humongous \humongous=0pt plus 1000pt minus 1000pt

\newif\ifdtup

\relax
\newcommand{\be}{\begin{equation}}
\newcommand{\ee}{\end{equation}}
\newcommand{\bea}{\begin{eqnarray}}
\newcommand{\eea}{\end{eqnarray}}
\newcommand{\pro}{\partial}
\newcommand{\n}{\hat n}
\newcommand{\oneg}{\displaystyle\frac{1}{g}}
\newcommand{\D}{{\hat D}}

\newcommand{\X}{{\vec X}}
\newcommand{\vX}{{\vec X}}

\newcommand{\A}{{\vec A}}

\newcommand{\hD}{{\hat D}}
\newcommand{\hn}{\hat n}

\newcommand{\tC}{{\tilde C}}

\newcommand{\dfrac}{\displaystyle\frac}
\newcommand{\ba}{\begin{array}}
\newcommand{\ea}{\end{array}}
\newcommand{\nn}{\nonumber}

\newcommand{\valpha}{{\vec \alpha}}

\begin{document}
\title{A Stable Magnetic Background in SU(2) QCD}
\bigskip
\author{Y.M. Cho}
\email{ymcho@yongmin.snu.ac.kr}
\affiliation{C. N. Yang Institute
for Theoretical Physics,
State University of New York, Stony Brook, New York 11794, USA \\
and \\
School of Physics, College of Natural Sciences, \\
Seoul  National University, Seoul 151-742, Korea}
\author{D.G. Pak}
\email{dmipak@phya.snu.ac.kr}
\affiliation{ Center for Theoretical Physics, 
Seoul National University, Seoul 151-742, Korea\\
and\\
Institute of Applied Physics, 
Uzbekistan National University, Tashkent 700-095, Uzbekistan}
                                                                                
\begin{abstract}
Motivated by the instability of the Savvidy-Nielsen-Olesen (SNO)
vacuum we make a systematic search for a stable magnetic background 
in pure $SU(2)$ QCD. It is shown that Wu-Yang monopole-antimonopole
pair is unstable under vacuum fluctuations. However, it is shown 
that a pair of axially symmetric monopole-antimonopole
string configuration is stable, provided the distance between the
two strings is small enough (less than a critical value).
The existence of a stable monopole-antimonopole string
background strongly supports that a magnetic condensation 
of monopole-antimonopole pairs can indeed generate 
a dynamical symmetry breaking, and thus
a desired magnetic confinement of color, in QCD.
                                                                                
\end{abstract}
                                                                                
\pacs{12.38.-t, 11.15.-q, 12.38.Aw, 11.10.Lm}
\keywords{stable magnetic background in QCD, magnetic confinement
in QCD}
\maketitle
                                                                                
\section{Introduction}
                                                                                
One of the most outstanding problems
in theoretical physics is the confinement problem in QCD. It has
long been argued that the monopole condensation could explain the
confinement of color through the dual Meissner effect \cite{nambu,cho1}.
Indeed, if one assumes the monopole condensation, one could easily argue that
the ensuing dual Meissner effect guarantees the confinement \cite{cho2}.
There have been many attempts to prove this scenario
in QCD \cite{savv,ditt}. Unfortunately the earlier attempts has failed to
establish the desired magnetic condensation,
because the magnetic background, known as the Savvidy-Nielsen-Olesen
(SNO) vacuum, is not stable.
In fact the effective action of QCD obtained with the SNO vaccum
develops an imaginary part, which implies that the SNO vacuum is
unstable \cite{niel,sch}.
This instability of the magnetic condensation
has been widely accepted and never been convincingly revoked.
                                                                                
In retrospect there are many reasons why the earlier attempts
have not been so successful. First, the calculation of
the effective action has involved tachyons which violates 
the causality, a fundamental principle in quantum field theory.
Indeed it is well-known that the imaginary part of 
the effective action originates from the tachyonic contribution. 
This tells that the causality principle may have been compromised in 
the calculation of the effective action. Secondly, the calculation of
the effective action was not gauge independent.
In fact the SNO background itself was not gauge 
invariant \cite{niel,sch}. And obviously any background
which is not gauge invariant can not possibly become
a stable vacuum. From these points of view it is really not
suprising that the SNO vacuum turns out to be unstable.
There have been attempts to cure this defect of 
the SNO vacuum and prove the magnetic condensation with a gauge 
invariant background, but unfortunately these attempts have not 
been very successful \cite{ditt,niel,sch}.
                                                                                
Recently, however, this instability of the SNO vacuum has been
studied more carefully. It has been shown that, if one uses 
a proper infra-red regularization which respects causality,
the imaginary part in the effective action
disappears \cite{cho3,cho99,cho4}.
Furthermore, it has been argued that the imaginary part 
of the effective action disappears if one imposes the
gauge invariance to the SNO vacuum correctly.
Indeed the calculation of the effective action
based on color reflection invariance shows that 
the effective action has no imaginary part \cite{qcd8}.
This implies that a ``gauge-invariant''
SNO vacuum can be qualified as  a stable vacuum of QCD.
                                                                                
Nevertheless, the meaning of the ``gauge-invariant''
SNO background has not been fully understood so far. 
In particular, an explicit example of stable magnetic background 
has never been constructed. To understand the complexity of the problem,
notice that the color charge in QCD can uniquely be defined
only after one selects the color direction. But the color direction
in QCD is gauge dependent. So it is a non-trivial matter
to construct a ``gauge-invariant''
magnetic vacuum.
                                                                                
There is an intuitive way to understand why
the SNO background is unstable. The SNO vacuum is a constant color
magnetic field pointed to a fixed direction in the color 
$SU(2)$ space. Unfortunately this vacuum configuration is 
not gauge invariant.
To see this consider a second magnetic vacuum
which has the opposite color direction.
Clearly the two vacua are distinct because they have
opposite color. Nevertheless they are gauge equivalent,
because one can always rotate the first vacuum to the second one
with a gauge transformation \cite{qcd8}. This means that neither
the first nor the second vacuum is gauge invariant.
Only the gauge invariant combination of the two vacua
becomes gauge invariant, and thus could become a physical vacuum.
This is why the SNO vacuum must be unstable.
                                                                                
There is another example which has exactly the same instability 
problem. Consider the Wu-Yang monopole \cite{wu,cho80}. 
In spite of its topological origin it is well known that
the monopole is unstable \cite{brandt}. This is because
it is not gauge invariant.
Here again we have the anti-monopole which is gauge equivalent
to the monopole, and only a gauge invariant combination 
of the monopole and anti-monopole
can exist as a physical (gauge invariant) object,
just as a gauge invariant
combination of quark and anti-quark can exist as physical.
This tells that a physical monopole condensation should not be
a simple monopole condensation,
but a condensation of gauge invariant combination of
monopole and anti-monopole pairs.
                                                                                
The above heuristic argument tells that only a gauge invariant
SNO vacuum, if at all, has a chance to become a physical
vacuum of QCD \cite{qcd8}. {\it The purpose of this paper is
to search for a stable magnetic background in $SU(2)$ QCD.
We analyze the stability of two classical magnetic
backgrounds, a pair of axially symmetric monopole-antimonopole 
strings and a pair of magnetic vortex-antivortex strings,  
and show that the pair of
monopole-antimonopole string configuration becomes stable
provided the distance between two strings is small enough.}
As far as we understand, the pair of axially symmetric 
monopole-antimonopole strings constitutes a first explicit example 
of a stable magnetic background in QCD. 
More importantly the result can serve as a strong argument that
a gauge invariant monopole-antimonopole condensation
can provide a stable vacuum in QCD.
This reinforces the claim that the ``gauge-invariant''
SNO vacuum could generates a desired dynamical symmetry breaking
which could confine the color in QCD.
                                                                                
The paper is organized as follows. In Section II we review
the geometric structure of the connection space in QCD,
and discuss how one can obtain a gauge independent
separation of a classical background from the quantum fluctuation. 
In Section III we review
the SNO effective action of QCD to clarify the origin of
the instability of the SNO vacuum.
In Section IV we discuss a gauge invariant calculation of 
the effective action, and show how the gauge invariance can
cure the instability of the SNO vacuum. 
In Section V we analyze the stability of the
classical Wu-Yang monopole and anti-monopole pair,
and show that the configuration is unstable.
In Section VI we consider an axially symmetric monopole string
(a two-dimensional classical monopole configuration), 
and show that it is unstable under the quantum fluctuation.
In Section VII we consider a pair of axially symmetric 
monopole and anti-monopole strings,
and show that the configuration is stable if the distance
between the two strings is small enough.
In Section VIII we study the stability of an axially symmetric 
magnetic vortex-antivortex pair, and show that the vortex pair 
is unstable. Finally in Section IX we discuss the physical 
significance of our result.
                                                                                
\section{Gauge Independent Decomposition of non-Abelian
Gauge Potential}
                                                                                
One of the conceptional problems in non-Abelian gauge theory
is how to define the color. It is well known that
the conserved color charge is gauge dependent. Indeed the
gauge-dependence of the conserved color is so severe that
one can always choose a gauge in which the color charge 
becomes identically zero. This means that, to
discuss the confinement of color, one must know how
to define the color in a gauge independent way.
Consider $SU(2)$ QCD for simplicity.  A natural way to define 
the color is to introduce an isotriplet (a unit vector field
in color space) $\n$ which selects the color direction
at each space-time point, and to
decompose the gauge potential into the restricted
potential $\hat A_\mu$ which leaves $\n$
invariant and the valence potential $\vec X_\mu$
which forms a covariant vector field \cite{cho1,cho2},
\bea
& \vec{A}_\mu =A_\mu \n - \oneg \n\times\pro_\mu\n+\X_\mu\nonumber
= \hat A_\mu + \X_\mu, \nn\\
&(\n^2 =1,~~~ \hat{n}\cdot\vec{X}_\mu=0), 
\label{cdec}
\eea
where $A_\mu = \n\cdot \vec A_\mu$
is the ``electric'' potential. Clearly this way of
selecting the color direction is gauge independent,
because $\n$ is chosen to be gauge covariant.
                                                                                
Notice that the restricted potential is precisely the connection which
leaves $\n$ invariant under the parallel transport,
\bea
\D_\mu \n = \pro_\mu \n + g {\hat A}_\mu \times \n = 0.
\eea
Under the infinitesimal gauge transformation
\bea
\delta \n = - \vec \alpha \times \n  \,,\,\,\,\,
\delta \A_\mu = \oneg  D_\mu \vec \alpha,
\eea
one has
\bea
&&\delta A_\mu = \oneg \n \cdot \pro_\mu \valpha,\,\,\,\
\delta \hat A_\mu = \oneg \D_\mu \valpha  ,  \nn \\
&&\hspace{1.2cm}\delta \X_\mu = - \valpha \times \X_\mu  .
\eea
This shows that $\hat A_\mu$ by itself describes
an $SU(2)$ connection which
enjoys the full $SU(2)$ gauge degrees of freedom. Furthermore
$\vec X_\mu$ transforms covariantly under the gauge transformation.
Most importantly, the decomposition is gauge-independent. Once
the color direction $\hn$ is selected, the
decomposition follows automatically,
independent of the choice of a gauge.
                                                                                
The restricted potential $\hat{A}_\mu$ actually has a dual structure.
Indeed the field strength made of the restricted potential is decomposed as
\begin{eqnarray}
&\hat{F}_{\mu\nu}=(F_{\mu\nu}+ H_{\mu\nu})\hat{n}\mbox{,}\nonumber\\
&F_{\mu\nu}=\partial_\mu A_{\nu}-\partial_{\nu}A_\mu, \nn\\
&H_{\mu\nu}=-\dfrac{1}{g} \hat{n}\cdot
(\partial_\mu\hat{n}\times\partial_\nu\hat{n})
=\partial_\mu \tilde C_\nu-\partial_\nu \tilde C_\mu,
\end{eqnarray}
where $\tilde C_\mu$ is the ``magnetic'' potential
\cite{cho1,cho2}.
Notice that we can always introduce the magnetic
potential (at least locally section-wise), because $H_{\mu\nu}$
forms a closed two-form
\bea
\partial_\mu {\tilde H}_{\mu\nu} = 0 ~~~~~~~ ( {\tilde H}_{\mu\nu} =
\dfrac{1}{2} \epsilon_{\mu\nu\rho\sigma} H_{\rho\sigma} ).
\eea
This allows us to  identify the non-Abelian magnetic potential by
\bea
\vec C_\mu= -\frac{1}{g}\hat n \times \partial_\mu\hat n.
\eea
Indeed with $\hn=\hat r$ the magnetic potential describes 
the well-known Wu-Yang minopole \cite{wu,cho80}.

With the decomposition (\ref{cdec}) one has
\bea
\vec{F}_{\mu\nu}&=&\hat F_{\mu \nu} + \D _\mu \vX_\nu -
\D_\nu \vX_\mu + g\vX_\mu \times \vX_\nu,
\eea
so that the Lagrangian can be written as follows
\bea
&{\cal L} =-\dfrac{1}{4} {\hat F}_{\mu\nu}^2 
-\dfrac{1}{4}(\D_\mu\vX_\nu-\D_\nu\vX_\mu)^2 \nn \\
&-\dfrac{g}{2} {\hat F}_{\mu\nu} \cdot (\vX_\mu \times \vX_\nu)
-\dfrac{g^2}{4} (\vX_\mu \times \vX_\nu)^2.
\eea
This shows that QCD is a restricted gauge theory which has
a gauge covariant valence gluon as a colored source.
                                                                                
The decomposition (\ref{cdec}), which has recently been referred to as
the ``Cho decomposition'' or ``Cho-Faddeev-Niemi
decomposition'' \cite{fadd,lang,zucc,kondo}, was first introduced
long time ago in an attempt to demonstrate
the monopole condensation in QCD \cite{cho1,cho2}.
But only recently the importance of the decomposition
in clarifying the non-Abelian dynamics
has become appreciated by many authors.
Indeed this decomposition has played a crucial role for us to
establish the Abelian dominance
in Wilson loops in QCD \cite{cho00}, and to clarify
the topological structure (in particular the Deligne 
cohomology) of the non-Abelian gauge theory \cite{zucc}.
                                                                                
An important advantage of the decomposition (1) is that it can actually
Abelianize (or more precisely ``dualize'') the non-Abelian
gauge theory \cite{cho1,cho2,cho01}. To see this let
$(\hat n_1,~\hat n_2,~\hat n)$ be a right-handed orthonormal 
basis in $SU(2)$ space and let
\begin{eqnarray}
&\vec{X}_\mu =X^1_\mu ~\hat{n}_1 + X^2_\mu ~\hat{n}_2\mbox{,} \nn\\
&(X^1_\mu = \hat {n}_1 \cdot \vec X_\mu,~~~X^2_\mu 
=\hat {n}_2 \cdot \vec X_\mu).            \nonumber
\end{eqnarray}
With this we have
\begin{eqnarray}
&\hat{D}_\mu \vec{X}_\nu =\Big[\partial_\mu X^1_\nu-g
(A_\mu+ \tilde C_\mu)X^2_\nu \Big]\hat n_1 \nn\\
&+ \Big[\partial_\mu X^2_\nu+ g (A_\mu+ \tilde C_\mu)X^1_\nu \Big]\hat{n}_2.
\end{eqnarray}
So introducing a dual
potential $B_\mu$ and a complex vector field $X_\mu$ by
\bea
& B_\mu = A_\mu + \tilde C_\mu , \nn\\
&X_\mu = \dfrac{1}{\sqrt{2}} ( X^1_\mu + i X^2_\mu ),
\eea
we can express the Lagrangian explicitly as follows,
\begin{eqnarray} 
\label{eq:Abelian}
&{\cal L}=-\dfrac{1}{4} G_{\mu\nu}^2
-\dfrac{1}{2}|\hat{D}_\mu{X}_\nu-\hat{D}_\nu{X}_\mu|^2
+ ig G_{\mu\nu} X_\mu^* X_\nu \nn\\
&-\dfrac{1}{2} g^2 \Big[(X_\mu^*X_\mu)^2-(X_\mu^*)^2 (X_\nu)^2 \Big] \nn\\
&= -\dfrac{1}{4}(G_{\mu\nu} + X_{\mu\nu})^2
-\dfrac{1}{2}|\hat{D}_\mu{X}_\nu-\hat{D}_\nu{X}_\mu|^2,
\end{eqnarray}
where
\bea
& G_{\mu\nu} = F_{\mu\nu} + H_{\mu\nu},
~~~\hat{D}_\mu{X}_\nu = (\partial_\mu + ig B_\mu) X_\nu, \nn\\
& X_{\mu\nu} = - i g ( X_\mu^* X_\nu - X_\nu^* X_\mu ). \nonumber
\eea
Clearly this describes an Abelian gauge theory coupled to
the charged vector field $X_\mu$.
But the important point here is that the Abelian potential
$B_\mu$ is given by the sum of the electric and magnetic potentials
$A_\mu+\tilde C_\mu$.
In this form the equations of motion of $SU(2)$ QCD is expressed by
\begin{eqnarray} \label{eom}
&\partial_\mu(G_{\mu\nu}+X_{\mu\nu}) = i g X^*_\mu
({\hat D}_\mu X_\nu -{\hat D}_\nu X_\mu ) \nn\\
&- i g X_\mu ({\hat D}_\mu X_\nu - {\hat D}_\nu X_\mu )^*, \nn\\
&\hat{D}_\mu(\hat{D}_\mu X_\nu- \hat{D}_\nu X_\mu)=ig X_\mu
(G_{\mu\nu} +X_{\mu\nu}).
\end{eqnarray}
This shows that one can indeed Abelianize the non-Abelian theory
with our decomposition. A remarkable feature of this Abelian
formulation is that here the topological field $\hat n$ is
replaced by the magnetic potential $\tilde C_\mu$ \cite{cho1,cho2}.
                                                                                
An important point of this Abelianization is that
it is gauge independent, because here we have never fixed
the gauge to obtain this Abelian formalism. So one might
ask how the non-Abelian gauge symmetry is realized in this Abelian
formalism. To discuss this let
\bea
&\vec \alpha = \alpha_1~\hn_1 + \alpha_2~\hn_2 + \theta~\hat n, \nn\\
&\alpha = \dfrac{1}{\sqrt 2} (\alpha_1 + i ~\alpha_2), \nn\\
&\vec C_\mu = - \dfrac {1}{g} \hn \times \partial_\mu \hn
= - C^1_\mu \hn_1 - C^2_\mu \hn_2, \nn\\
&C_\mu = \dfrac{1}{\sqrt 2} (C^1_\mu + i ~ C^2_\mu).
\eea
Then the Lagrangian (\ref{eq:Abelian}) is invariant not only under
the active gauge transformation (4) described by
\bea \label{eq:active}
&\delta A_\mu = \dfrac{1}{g} \partial_\mu \theta -
i (C_\mu^* \alpha - C_\mu \alpha^*),
~~~&\delta \tilde C_\mu = - \delta A_\mu, \nn\\
&\delta X_\mu = 0,
\eea
but also under the following passive gauge transformation
described by
\bea 
\label{eq:passive}
&\delta A_\mu = \dfrac{1}{g} \partial_\mu \theta -
i (X_\mu^* \alpha - X_\mu \alpha^*), ~~~&\delta \tilde C_\mu = 0, \nn\\
&\delta X_\mu = \oneg \hD_\mu \alpha - i \theta X_\mu.
\eea
Clearly this passive gauge transformation assures the desired
non-Abelian gauge symmetry in the Abelian formalism.
This tells that the Abelian theory not only retains
the original gauge symmetry, but actually has an enlarged (both the
active and passive) gauge symmetries.
But we emphasize that this is not the ``naive'' Abelianization
of QCD which one obtains by fixing the gauge.
Our Abelianization is a gauge-independent Abelianization.
                                                                                
\section{Savvidy-Nielsen-Olesen Effective Action: A Review}
                                                                                
To calculate the one-loop effective action one must divide
the gluon field into two parts, the slow-varying
classical background $\vec B_\mu$ and the fluctuating quantum
part $\vec Q_\mu$,
\bea
\vec A_\mu = \vec B_\mu + \vec Q_\mu,
\label{d}
\eea
and integrate the quantum part \cite{dewitt,pesk}.
Of course, the separation of the quantum part
from the classical background
has to be gauge independent for the effective action to be
gauge independent. The decomposition (\ref{cdec}) 
is very useful for this purpose, because it naturally provides 
the gauge independent separation of the classical 
background from the quantum fluctuation.
Indeed the gauge independent separation follows automatically
if we identify the classical background to be 
the restricted potential $\hat A_\mu$ and the quantum 
fluctuation to be the valence potential $\vec X_\mu$. 

In the Abelian formalism this means that we can treat 
$B_\mu$ as the classical background and $X_\mu$ as 
the fluctuating quantum part.
In this picture the active gauge transformation 
(\ref{eq:active}) is viewed as the background gauge
transformation and the passive gauge transformation (\ref{eq:passive}) 
is viewed as the quantum gauge transformation.
To calculate the one-loop effective action, 
we fix the gauge of the quantum gauge transformation by
imposing the following gauge condition to $X_\mu$,
\bea
&\hat D_\mu X_\mu =0, ~~~~~(\hat{D}_\mu X_\mu)^* =0 \nn\\
&{\cal L}_{gf} =- \dfrac{1}{\xi}
|{\hat D}_\mu X_\mu|^2.
\eea
Under the gauge transformation (\ref{eq:passive})
the gauge condition
depends only on $\alpha$, so the corresponding
Faddeev-Popov determinant is given by
\be \label{eq:complexFP}
M_{FP} = \left| \begin{array}{cc}
\dfrac{\delta (\hat{D}_\mu X_\mu)}{\delta \alpha} &
\dfrac{\delta (\hat{D}_\mu X_\mu)}{\delta \alpha^*} \\
\dfrac{\delta (\hat{D}_\mu X_\mu)^*}{\delta \alpha} &
\dfrac{\delta (\hat{D}_\mu X_\mu)^*}{\delta \alpha^*}
\end{array} \right|.
\ee
With this gauge fixing
the one-loop effective action takes the following 
form \cite{ditt,niel,cho3,cho4},
\begin{widetext}
\bea \label{eq:effaction}
&\exp \Big[iS_{eff}(B_\mu) \Big] = \dfrac{}{} \int
{\cal D} X_\mu {\cal D} X_\mu^* {\cal D}c_1{\cal D}c_1^{\dagger}
{\cal D}c_2{\cal D}c_2^{\dagger} \exp \Big{\lbrace} \dfrac{}{}
i\int \Big[ -\dfrac{1}{4}(G_{\mu\nu} + X_{\mu\nu})^2
-\dfrac{1}{2}|\hat{D}_\mu{X}_\nu-\hat{D}_\nu{X}_\mu|^2 \nn\\
&-\dfrac{1}{\xi} |\hat {D}_\mu X_\mu|^2
+ c_1^\dagger (\hat{D}^2
+ g^2X_\mu^* X_\mu )c_1 - g^2 c_1^\dagger
 X_\mu X_\mu c_2 + c_2^{\dagger}
(\hat{D}^2 + g^2X_\mu^* X_\mu )^*c_2
- g^2 c_2^{\dagger} X^*_\mu X^*_\mu c_1 ~\Big] d^4x \Big{\rbrace},
\eea
where $c_1$ and $c_2$ are the complex ghost fields.
To evaluate the integral we
notice that the functional determinants of the valence gluon
and the ghost loops are expressed as
\bea
&{\rm Det}^{-\frac{1}{2}} K_{\mu \nu}\simeq
{\rm Det}[-g_{\mu \nu}
 (\hat D \hat D)+ 2ig G_{\mu \nu}],
~~~~~~~{\rm Det} M_{FP} = {\rm Det} [-(\hat D \hat D)]^2.
\eea
Using the relation
\bea
&G_{\mu \alpha} G_{\nu \beta} G_{\alpha \beta} = \dfrac{1}{2} G^2
G_{\mu \nu} +\dfrac{1}{2}(G \tilde G) {\tilde G}_{\mu \nu},
~~~~~~~({\tilde G}_{\mu \nu}=\dfrac{1}{2}{\epsilon}_{\mu\nu\rho\sigma}
G_{\rho\sigma}),
\eea
we can simplify the functional determinants of the gluon
and the ghost loops as follows,
\bea \label{eq:functDet}
& \ln {\rm Det}^{-\frac 1 2} K = \ln {\rm Det} [(-\hD^2+2a)(-\hD^2-2a)]
+ \ln {\rm Det} [(-\hD^2-2ib)(-\hD^2+2ib)],\nn\\
& \ln {\rm Det}M_{FP} = 2\ln {\rm Det}(-\hD^2),
\eea
where
\bea
a = \dfrac{g}{2} \sqrt {\sqrt {G^4 + (G \tilde G)^2} + G^2},
~~~~~~~b = \dfrac{g}{2} \sqrt {\sqrt {G^4 + (G \tilde G)^2} - G^2}. \nn
\eea
\end{widetext}
Notice that two determinants ${\rm Det}(-\hD^2 \pm 2a)$ 
(and ${\rm Det}(-\hD^2 \pm 2ib)$)
correspond to two spin orientations of the valence gluon.

Savvidy has chosen a covariantly constant color magnetic field
as the classical background \cite{savv,niel,ditt}
\bea
&\vec B_\mu = \dfrac{1}{2} H_{\mu\nu} x_\nu \n_0,
~~~~~\vec G_{\mu\nu} = H_{\mu\nu} \n_0, \nn\\
&\bar D_\mu \vec G_{\mu\nu} = 0,
\label{sb}
\eea
where $H_{\mu\nu}$ is a constant magnetic field and $\n_0$ is
a constant unit isovector.
With this one has
\bea
&\Delta S = i \ln {\rm Det} [(-\hD^2+2gH)(-\hD^2-2gH)] \nn\\
&- 2i \ln {\rm Det}(-\hD^2), \nn\\
&H=\sqrt {\dfrac{H_{\mu\nu}^2}{2}}.
\label{det}
\eea
One can evaluate the functional determinants using
Schwinger's proper time method \cite{schw}, and find
\bea
&\Delta{\cal L} = \dfrac{1}{16 \pi^2}\int_{0}^{\infty}
\dfrac{dt}{t^2} \dfrac{gH/ \mu^2}{\sinh (gHt/\mu^2)}
\Big[\exp (-2gHt/\mu^2 ) \nn\\
&+  \exp (+2gHt/\mu^2) \Big],
\label{eam}
\eea
where $\mu$ is a dimensional parameter. The integral has a severe
infra-red divergence, and to perform the integral
we have to regularize the infra-red
divergence first. Let us regularize it with the $\zeta$-function
regularization. From the definition of the generalized
$\zeta$-function \cite{table}
\bea
&\zeta (s,\lambda) = \dfrac{}{}\sum_{n=0}^{\infty}
\dfrac{1}{(n+\lambda)^s} \nn\\
&= \dfrac{1}{\Gamma(s)} \int_0^{\infty} \dfrac{x^{s-1} \exp(-\lambda x)}
{1-\exp(-x)} dx,
\label{zeta}
\eea
we have
\begin{widetext}
\bea
&\Delta {\cal L} = \dfrac{}{} \lim_{\epsilon \rightarrow 0}
\dfrac{1}{16 \pi^2} \int_{0}^{\infty} \dfrac{dt}{t^{2-\epsilon}}
\dfrac{gH \mu^2}{\sinh (gHt/\mu^2)}
\Big[\exp (-2gHt/\mu^2) + \exp (+2gHt/\mu^2) \Big] \nn\\
&= \dfrac{}{} \lim_{\epsilon \rightarrow 0}
\dfrac{gH \mu^2}{8 \pi^2} \int_{0}^{\infty} \dfrac{dt}{t^{2-\epsilon}}
\dfrac{\exp (-3gHt/\mu^2) + \exp (+gHt/\mu^2)}{1-\exp(-2gHt/\mu^2)} \nn\\
&= \dfrac{}{} \lim_{\epsilon \rightarrow 0}
\dfrac{g^2H^2}{4 \pi^2} (\dfrac{2gH}{\mu^2})^{-\epsilon} \Gamma(\epsilon-1)
\Big[\zeta(\epsilon-1,\dfrac{3}{2}) + \zeta(\epsilon-1,-\dfrac{1}{2})\Big]\nn\\
&= \dfrac{}{} \lim_{\epsilon \rightarrow 0}
\dfrac{g^2H^2}{4 \pi^2} (1-\epsilon \ln \dfrac{2gH}{\mu^2})
\big(\dfrac{1}{\epsilon} -\gamma +1 \big)
\Big[\big(\zeta(-1,\dfrac{3}{2}) + \zeta(-1,-\dfrac{1}{2})\big)
+ \epsilon \big(\zeta'(-1,\dfrac{3}{2})
+ \zeta'(-1,-\dfrac{1}{2})\big)\Big] \nn\\
&= \dfrac{11 g^2H^2}{48 \pi^2} \big(\dfrac{1}{\epsilon} - \gamma
+1 - \ln \dfrac{2gH}{\mu^2} \big)
- \dfrac{g^2H^2}{4 \pi^2} \big(2 \zeta'(-1,\dfrac{3}{2})
- i \dfrac{\pi}{2}\big),
\label{zetareg}
\eea
where $\zeta' = \dfrac{d\zeta}{ds} (s,\lambda)$,
and we have used  the fact \cite{table}
\bea
&\zeta(-1,\dfrac{3}{2}) = \zeta(-1,-\dfrac{1}{2}) = -\dfrac{11}{24},
~~~~~\zeta'(-1,-\dfrac{1}{2}) = \zeta'(-1,\dfrac{3}{2}) 
- i \dfrac{\pi}{2}. \nn
\eea
So, with the ultra-violet regularization by modified
minimal subtraction we obtain the SNO effective action
\bea
&{\cal L}_{eff}=-\dfrac{H^2}{2} -\dfrac{11g^2}{48\pi^2}H^2(\ln
\dfrac{gH}{\mu^2}-c) 
+ i \dfrac {g^2} {8\pi} H^2, \nn\\
&c=1-\ln 2 -\dfrac {24}{11} \zeta'(-1, \frac{3}{2})=0.94556....
\label{snoea}
\eea
\end{widetext}
The real part of the effective action has a non-trivial SNO
vacuum at $<H> \neq 0$. Unfortunately the effective action
contains the well-known imaginary part which destabilzes the SNO
vacuum.
                                                                                
\begin{figure*}
\includegraphics{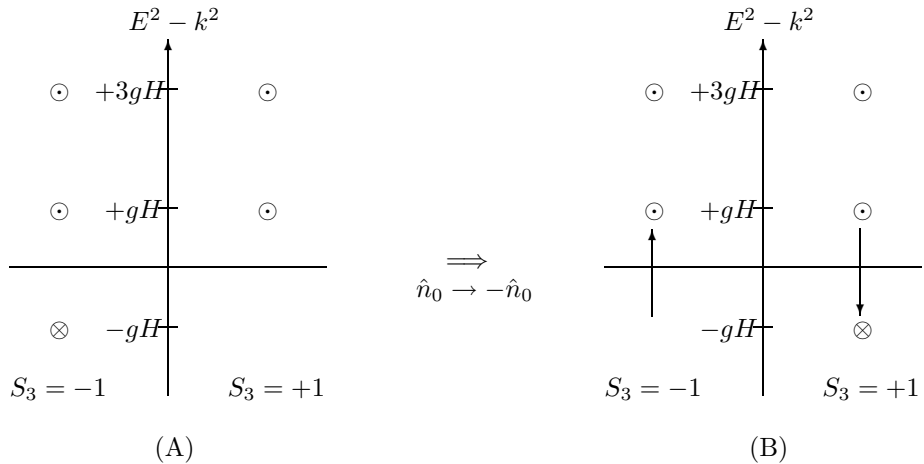}
\caption{\label{Fig. 1} The eigenvalues of the functional
determinant of the gluon loop. When the gluon spin is
anti-parallel to the magnetic field ($S_3=-1$), the ground state
(with $n=0$) becomes tachyonic when $k^2<gH$. Notice, however,
that under the color reflection of $\hn_0$ to $-\hn_0$
$H$ changes to $-H$ so that the eigenvalues change from (A)
to (B). This shows that the spin polarization direction
of gluon with respect to the magnetic field is a gauge artifact. 
This excludes the tachyons from
the functional determinant.}
\end{figure*}

\section{Gauge Invariant Calculation of Effective Action}

Notice that the imaginary part of the effective action originates 
from the determinant ${\rm Det}(-\hat D^2-2gH)$, which corresponds 
to the gluon loop whose spin is anti-parallel to the magnetic 
field \cite{niel}. Indeed in the Abelian formalism 
the calculation of the functional
determinants ${\rm Det}(-\hat D^2 \pm 2gH)$ in (\ref{det}) 
amounts to the calculation of
the energy eigenvalues of a massless charged vector field
(the valence gluon) in a constant external
magnetic field $H_{\mu\nu}$.
Choosing the direction of the magnetic field to be the $z$-direction,
one obtains the well-known eigenvalues
\bea
&E = 2gH (n + \dfrac{1}{2}) + k^2 \pm 2gH, \nn\\
&H = H_{12},
\label{ev}
\eea
where $k$ is the momentum of the eigen-function in $z$-direction.
Notice that the $\pm$ signature correspond to the spin $S_3=\pm 1$
of the valence gluon. So, when $n=0$,
the eigen-functions with $S_3=-1$ have an imaginary energy when
$k^2<gH$, and thus become tachyons which violate
the causality. And these tachyonic eigenstates create 
the instability of the magnetic background and the imaginary 
part in the effective action \cite{niel,cho3,cho4}.  

The existence of the tachyonic modes in the functional determinant
tells that the calculation of the effective action violates
the causality. This means that we must exclude the unphysical
tachyonic modes from the functional determinant. 
The question is how. A natural answer is to impose 
the causality in the calculation of the effective 
action \cite{cho3,cho4}.
To show this we go to the Minkowski time with
the Wick rotation, and find that (\ref{eam}) changes to
\bea
&\Delta{\cal L} = -\dfrac{1}{16 \pi^2}\int_{0}^{\infty}
\dfrac{dt}{t^2} \dfrac{gH/ \mu^2}{\sin (gHt/\mu^2)}
\Big[\exp (-i2gHt/\mu^2 ) \nn\\
&+  \exp (+i2gHt/\mu^2) \Big],
\label{eam1}
\eea
In this form the infra-red divergence has disappeared,
but now we face an ambiguity in choosing the correct contours
of the integrals in (\ref{eam1}). But this ambiguity can
be resolved by causality. Indeed
the standard causality argument requires us to
identify $2gH$ in the first integral as
$2gH-i\epsilon$, but in the second integral as
$2gH+i\epsilon$. This tells that the poles in the first integral 
in (\ref{eam1}) should lie above
the real axis, but the poles in the second integral should lie
below the real axis \cite{cho3,cho4}. With this causality 
requirement the two integrals
become complex conjugate to each other. This tells that
the effective action must be explicitly real, 
without any imaginary part.

The fact that the causality removes the tachyonic modes 
in the functional determinant is perhaps not surprising. 
What is surprizing is that a completely independent principle,
the gauge invariance, can also remove the tachyonic modes \cite{qcd8}.
To see this, notice that the Savvidy background (\ref{sb}) is
not gauge invariant. Indeed $\vec G_{\mu\nu}$
must be gauge covariant. So one can
change $\vec G_{\mu\nu}$ to $-\vec G_{\mu\nu}$, and thus
$H_{\mu\nu}$ to $-H_{\mu\nu}$, by a gauge transformation
(with the color reflection of $\hn_0$ to $-\hn_0$). 
In this case ${\rm Det} (-\hat D^2 + 2gH)$ changes 
to ${\rm Det} (-\hat D^2-2gH)$, and vise versa, under the gauge 
transformation. On the other hand the gauge transformation
does not affect the gluon spin. This means that one can change
the direction of magnetic field with respect to
the spin polarization direction of gluon by a gauge transformation.
In other words the spin polarization direction of the gluon
with respect to the magnetic
background is a gauge artifact. More importantly 
the eigenvalues of the $S_3=+1$ gluon shift nagatively,
and those of the $S_3=-1$ gluon shift positively, by a factor $2gH$ 
under the gauge transformation.
And obviously only the eigenvalues
which are invariant under this transformation should qualify
to be gauge invariant. This means that the gauge invariant
eigenstates are those
which are independent of the spin orientation
of the valence gluon which appear in both $S_3=+1$
and $S_3=-1$ simultaneously. 

This is shown schematically in Fig. 1,
where (A) transforms to (B) under the color reflection. 
This clearly tells that the tachyonic modes which caused 
the instability of the SNO vacuum is not gauge invariant, 
and thus should not be included in the calculation of 
the gauge-invariant functional determinants. This means that 
a gauge invariant calculation of the effective action must
produce a stable magnetic condensation. 

One might think (incorrectly) that the eigenvalues of the functional
determinants do not change because ${\rm Det}(-\hat D^2 \pm 2gH)$ 
remain unchanged under the gauge transformation, 
even though ${\rm Det} (-\hat D^2+2gH)$ changes
to ${\rm Det} (-\hat D^2-2gH)$ and vise versa. 
This is wrong, because here one must calculate the eigenvalues 
of the spin-up gluon and spin-down gluon separately.
This is very important. And the determinant for the spin-up gluon 
${\rm Det} (-\hat D^2+2gH)$ changes to ${\rm Det} (-\hat D^2-2gH)$
and the determinant for the spin-down gluon ${\rm Det} (-\hat D^2-2gH)$
changes to ${\rm Det} (-\hat D^2+2gH)$. So the eigenvalues change, and
for each spin polarization only the positive eigenvalues remain
invariant under the gauge transformation.

With the gauge invariant calculation of the the functional
determinants, the effective action (\ref{eam}) changes to
\bea
&\Delta{\cal L} = \dfrac{1}{16 \pi^2}\int_{0}^{\infty}
\dfrac{dt}{t^2} \dfrac{gH/ \mu^2}{\sinh (gHt/\mu^2)}
\Big[\exp (-2gHt/\mu^2 ) \nn\\
&+ \exp (-2gHt/\mu^2) \Big],
\label{eam2}
\eea
which has no infra-red divergence at all.
This precludes the necessity to make any infra-red regularization.
>From this we have \cite{cho3,cho4,qcd8}
\bea
&{\cal L}_{eff} = -\dfrac{H^2}{2} -\dfrac{11g^2}{96\pi^2} H^2 (\ln
\dfrac{gH}{\mu^2}-c),
\label{eaho}
\eea
which clearly has no imaginary part.

A physical way to understand the above result is to remember
that the one-loop effective action is nothing but the vacuum to vacuum
amplitude in the presence of the classical background,
\bea
&\exp \Big[i S_{eff} (\vec B_\mu)\Big]
= <\Omega_+|~\Omega_-> \Big|_{\vec B_\mu} \nn\\
&= \dfrac{}{} \sum_{|n_i>}<\Omega_+|~n_i>
<n_i~|~\Omega_-> \Big|_{\vec B_\mu},
\label{vtv}
\eea
where $|\Omega>$ is the vacuum and $|n_i>$ is
a complete set of orthonormal states
of QCD. In this picture the gluon loop integral corresponds to
the summation of the complete set of states.
And obviously the complete set should not include
the tachyons, unless one wants to assert that the physical
spectrum of QCD must contain the unphysical tachyons
which violate (not only the causality but also) the gauge invariance.
This justifies the exclusion of the tachyons
in the calculation of the functional determinant.

The above discussion tells that $SU(2)$ QCD is able to generate
a stable magnetic condensation. But the stability of
the Savvidy vacuum has been so controversal that one might
like to see more evidence to support the stable magnetic condensation.
A best way to do this is to construct a classical magnetic background
which has no unstable modes. In the following we will
show that an axially symmetric monopole string and anti-monopole 
string pair has no unstable modes under the quantum fluctuation,
if the distance between two strings is small enough,
smaller than a critical value.

\section{Instability of Monopole-Antimonopole Background}
                                                                                
Before we discuss the stability of monopole-antimonopole 
pair, we first review the instability of the Wu-Yang 
monopole because two problems are closely related \cite{brandt}. 
In our notation the Wu-Yang monopole 
of charge $q/g$ is described by \cite{cho80} 
\bea
&\vec A_\mu=-\dfrac{1}{g} \n \times \pro_\mu \n, \nn\\
&\n = \Bigg(\matrix{\sin \theta \cos{q\phi} \cr
\sin \theta \sin{q\phi} \cr 
\cos \theta}\Bigg), 
\label{mono}
\eea
where $(r,\theta,\phi)$ is the spherical coordinates
and $q$ is an integer.
But in the Abelian formalism it is more convenient for us 
to describe the monopole in terms of the magnetic potential
$\tilde C_\mu$,
\bea
&\tC_\mu = \dfrac{q}{g} (\cos \theta -1) \partial_\mu \phi, 
\label{paramcho}
\eea
which is nothing but the Dirac's monopole potential.

To study the stability of the monopole background
we consider the functional determinant of the monopole 
which provides the one-loop correction to the effective action
\bea
&{\rm Det} \, K = {\rm Det}(-\tilde D^2 \pm 2\dfrac{q}{r^2}).
\eea
Notice that the functional determinant is precisely the one
we introduced in (\ref{det}), exept that here $H$ is given by 
the magnetic field strength of the monopole. 

Now, the absence or presence of negative modes of the operator $K$
implies stability or instability of the classical
background against small fluctuations of the gauge potential. 
To calculate the eigenvalues of the operator $K$ one can rewrite 
the eigenvalue equation as the following Schroedinger type equation
acting on a complex scalar field $\Psi$,
\bea
&K \Psi(r,\theta, \phi) = E \Psi(r, \theta, \phi),  \nn\\
&K = -\Delta - 2\dfrac{iq}{r^2\sin^2 \theta} \cos \theta \pro_\phi
+ \dfrac{q^2}{r^2} \cot^2 \theta \pm 2\dfrac{q}{r^2}, \nn \\
&\Delta = \dfrac{1}{r^2}\pro_r( r^2 \pro_r) + \dfrac{1}{r^2 \sin \theta}
\pro_\theta (\sin \theta \pro_\theta) + \dfrac{1}{r^2 \sin^2 \theta}
\pro^2_\phi \nn \\
&=\dfrac{1}{r^2}\pro_r( r^2 \pro_r) - \dfrac{\hat L^2}{r^2}, \nn\\
&\hat L^2 = - \big(\dfrac{1}{\sin \theta}
\pro_\theta (\sin \theta \pro_\theta) + \dfrac{1}{\sin^2 \theta}
\pro^2_\phi \big),
\label{schr}
\eea
where $\hat L$ is the angular momentum operator. 
Notice that here again the $\pm$ signatures represent
two spin orientations of the valence gluon.            
                                                                    
With 
\bea
\Psi(r, \theta, \phi)= R(r) Y(\theta,\phi),
\eea
one obtains the equation for the angular eigenfunction 
$Y(\theta,\phi)$ from (\ref{schr})
\bea
&\Big (\hat L^2 - \dfrac{2iq\cos \theta}{\sin^2 \theta} \pro_\phi 
+ q^2 \cot^2 \theta  \Big) Y(\theta,\phi)  \nn\\
&= \lambda Y(\theta, \phi).
\label{mhe}
\eea
Moreover, with
\bea
& Y(\theta,\phi) =  \dfrac{}{}\sum_{m=-\infty}^{+\infty}
\Theta_m (\theta) \Phi_m (\phi), \nn \\
& \Phi_m (\phi) = \dfrac{1}{\sqrt {2\pi}} \exp(im\phi).
\eea
one can reduce (\ref{mhe}) to
\bea
\Big(-\dfrac{1}{\sin \theta} \pro_\theta (\sin \theta \pro_\theta) 
+ \dfrac{(m +q \cos \theta)^2 }{\sin^2 \theta} \Big)\Theta
= \lambda \Theta.
\eea
This is exactly the eigenvalue equation for the monopole
harmonics which has been well-studied in the literature \cite{wu}.
>From the equation one obtains the following expression 
for the monopole harmonics and the corresponding 
eigenvalue spectrum 
\bea
& Y_{qjm}(\theta,\phi) =
\Theta_{qjm}(\theta) \Phi_m (\phi), \nn \\
&\Theta_{qjm}(\theta) = (1-\cos \theta)^{\gamma_+}
(1+ \cos \theta)^{\gamma_-}  \nn\\
&\times P_k (\cos \theta), \nn\\
&\lambda = (k+p)(k+p+1) - q^2 = j(j +1) -q^2, \nn \\
& p =\gamma_+ + \gamma_-,
~~~~\gamma_+=\dfrac{|m+q|}{2},~~~~\gamma_-=\dfrac{|m-q|}{2}, \nn\\
&j = k+p = k + \dfrac{1}{2} (|m-q| + |m+q|), \nn\\
&k=0,1,2,..., 
\label{qnumber}
\eea
where $P_k(x)$ is the Legendre polynomial of order $k$.
The quantum number $j$ is analogous to the orbital 
angular momentum quantum number $l$ of the standard spherical 
harmonics $Y_{lm}$, except that here $j$ starts from 
a non-zero integer value for a non-vanishing monopole 
charge $q$. 
                                                                                
Now, consider the equation for the radial
eigenfunction
\bea
&\Big(\dfrac{1}{r^2} \dfrac{d}{dr} ( r^2 \dfrac{d}{dr}) 
-\dfrac {1}{r^2} \big[j(j +1) -q^2\big] 
\mp 2\dfrac{q}{r^2} +E \Big)R(r) \nn \\
&= 0.
\eea
With $R(r)=\dfrac{1}{r}\chi(r)$ one obtains
\bea
&\Big(\dfrac{d^2}{dr^2} - \dfrac{1}{r^2}\big[j(j+1) - q^2 \pm 2 q \big]
+ E \Big) \chi(r) \nn\\
&=0. 
\label{eqnchi}
\eea
The equation has a general solution in terms of Bessel functions
of the first kind $J_\nu (z)$
\bea
&\chi (r) = \sqrt r \Big[C_1 J_{-\nu} (\sqrt E r) + C_2
J_\nu (\sqrt E r) \Big], \nn \\
& \nu = \dfrac{1}{2} \sqrt {1 + 4 [j(j+1) - q^2 \pm 2 q]},
\eea
where $C_i~(i=1,2)$ are the integration constants.
For positive values of $\nu$ and $E$ the finite solutions
oscillating at the infinity and vanishing at the origin
are given by $C_1 =0$.
The negative eigenvalues of $E$ can come
only from (\ref{eqnchi})
with the lower negative sign (which corresponds
to the operator $-\tilde D^2 - 2q/r^2$) and the lowest value of
$j=1$ with $q=1$. In this case we have to solve the equation
\bea
\Big(\dfrac{d^2}{dr^2} + \dfrac{1}{r^2} + E \Big) \chi =0,
\label{chi}
\eea
which is nothing but the one-dimensional 
Schroedinger equation for the attractive
potential $-1/r^2$. So the the monopole stability problem
is reduced to the well known one-dimensional 
quantum mechanical problem with an attractive potential 
proportional to $-1/r^2$ \cite{landau}.

With $\chi (r) = \sqrt r y(r)$ and $z = \sqrt E r$, 
(\ref{chi}) reduces to
\bea
\Big(z^2 \dfrac{d^2}{dz^2} + z \dfrac{d}{dz} + (z^2 + \dfrac{3}{4}) 
\Big)y =0.
\eea
In this form the energy dependence of the equation disappears 
completely, so that we have no condition of discreteness for 
the eigenvalue spectrum. The analytic solution to the equation 
is given by the Bessel function of the first kind
\bea
y=J_{i \sqrt {3/4}} (z).
\eea 
For the real argument $z$ 
(i.e. for the positive energies $E$), the real and imaginary
parts of the solution provide two independent
solutions vanishing at the infinity
and oscillating at the origin.
For negative energies (i.e. for the pure imaginary $z$), 
the real part of the Bessel function diverges at the infinity,
so that only the imaginary part of the
Bessel function becomes a physical solution. This solution
has a continuum of negative eigenvalue spectrum, 
and is oscillating near the origin and approaches to zero 
exponentially at the infinity. Both solutions, for
positive and negative energies, the radial eigenfunction $R(r)$ 
behaves like 
\bea
R(r)\simeq \dfrac{\sin \log (\sqrt {|E|} r)+ {\rm const}}{\sqrt r},
\eea 
near the origin. The solutions have
infinite number of zeros approaching the point $r=0$, 
so that for the negative energies
the valence gluon moving around the monopole must fall down 
to the center \cite{brandt}. 
For higher monopole charges
the qualitative picture remains the same, and one still has 
a continious bound state energy spectrum.
                                                                                
The above analysis implies that the undesired attractive 
potential proportional to $-1/r^2$ in (\ref{eqnchi}) vanishes 
when $q=0$, in which case $j$ starts from zero.
This can serve as a hint that one might expect the absence of
negative modes only for a magnetic background  
with vanishing monopole charge. So it is important to check 
the stability of magnetic background with vanishing monopole charge.

A simplest magnetic background which has a vanishing monopole charge
is a monopole-antimonopole pair. But in the following we
show that the monopole-antimonopole background is not
stable. Consider a monopole-antimonopole background
where the monopole and anti-monopole are located at 
the point $(x=y=0,z=a)$ and at the origin $(x=y=z=0)$.
The magnetic fields of the monopole and anti-monopole are
expressed by
\bea
& \vec H_+ = \dfrac{\vec r-\vec a}{|\vec r-\vec a|^3},
~~~~~\vec H_- = -\dfrac{\vec r}{r^3}, \nn \\
&\vec H = \vec H_+ +\vec H_-, \nn \\
&H = \dfrac{1}{r^2} \Big(1+ \dfrac{r^4}{r'^4} 
-2\dfrac{r^3}{r'^3}(1-\dfrac{a}{r}\cos \theta) \Big)^{\frac{1}{2}} \nn\\
&r'=r(1-2\dfrac{a}{r}\cos \theta+\dfrac{a^2}{r^2})^{\frac{1}{2}}.
\eea
The corresponding magnetic potential $\tC_\mu$ is given by
\bea
& \tC_\mu =\Big[(1-\dfrac{r}{r'})\cos \theta +\dfrac{a}{r'} \Big] 
\pro_\mu \phi.
\eea
With the magnetic background we can repeat the above stability 
analysis to see whether the monopole-antimonopole pair is stable
or not.                             
                                                     
The above background gives the following eigenvalue equation 
similar to (\ref{schr})
\bea
&\Big(\dfrac{1}{r^2}\pro_r( r^2 \pro_r) + \dfrac{1}{r^2 \sin \theta} 
\pro_\theta (\sin \theta \pro_ \theta) 
+ \dfrac{1}{r^2 \sin^2 \theta} \pro_\phi^2  \nn \\
&+\dfrac{2i}{r^2 \sin^2 \theta} \big[(1 - \dfrac{r}{r'}) \cos \theta 
+\dfrac{a}{r'} \big] \pro_\phi  \nn\\
&- \dfrac{1}{r^2 \sin^2 \theta} \big[(1 - \dfrac{r}{r'}) \cos \theta 
+\dfrac{a}{r'} \big]^2 \mp 2 H + E \Big) \Psi \nn\\
&=0.
\eea
Here again the $\mp H$ term correspond to two spin orientations
of the valence gluon with respect to the magnetic field.  
Due to the axial symmetry we can put 
\bea
\Psi(r,\theta,\phi) =\dfrac{}{}\sum_{m=-\infty}^{+\infty}
F_m(r,\theta) \Phi_m(\phi),
\eea
and obtain the equation for $F_m(r,\theta)$
\bea
&&\Big(\dfrac{1}{r^2}\pro_r(r^2 \pro_r) 
+ \dfrac{1}{r^2 \sin \theta} \pro_\theta 
(\sin \theta \pro_ \theta) -U + E \Big) F =0 , \nn \\
&&U = \dfrac{1}{r^2 \sin^2 \theta} \big[(1 - \dfrac{r}{r'}) 
\cos \theta + \dfrac{a}{r'} +m \big]^2 \pm 2H.
\eea
To find the energy spectrum let us consider the equation 
near the origin. Keeping the leading terms in the limit
$r$ goes to zero we can approximate
the equation as
\bea
&\Big(\dfrac{1}{r^2}\pro_r( r^2 \pro_r) 
+ \dfrac{1}{r^2 \sin \theta} \pro_\theta (\sin \theta \pro_\theta) \nn \\
&-\dfrac{(m+1 + \cos \theta )^2}{r^2 \sin^2 \theta}
\mp \dfrac{2}{r^2} + E \Big) F \simeq 0.
\eea
The equation is separable, and the angular part
of the eigenfunction is expressed by the monopole harmonic
function $\Theta_{1jm}$ which satisfies the
eigenvalue equation
\bea
&\Big(\dfrac{1}{\sin \theta} \pro_\theta
(\sin \theta \pro_\theta) - \dfrac{(m+1 + \cos \theta)^2}{\sin^2 \theta} \nn\\
& + \big[j(j +1) -1 \big] \Big) \Theta_{1jm} = 0,
\eea
where $j$ is given by
\bea
j = \dfrac{|m|}{2} + \dfrac{|m+2|}{2}.
\eea
Clearly the lowest value of $j$ is $1$,
which tells that the radial part of the potential $U$
still has an attractive potential proportional to $-1/r^2$.
>From this we conclude that
the monopole-antimonopole pair has to be unstable.
                                                                                
The lesson from this analysis is clear. 
Although the total magnetic charge of monopole-antimonopole pair
is zero, the potential of the pair near the origin still retains
the undesirable attractive potential which makes the magnetic background 
unstable. 

\section{Axially Symmetric Monopole String Background}
                                                                                
Now we consider the axially symmetric monopole string,
which can be regarded as an infinite string carrying homogeneous 
monopole charge density along the string.
The magnetic field strength of the axially symmetric
monopole string can be written in a simple form in the cylindrical
coordinates $(\rho,\phi,z)$
\bea
&\vec H = \dfrac {\alpha}{\rho} \hat \rho, \nn \\
&\tC_\mu = -\alpha (z + \tau) \pro_\mu \phi,
\label{mstring}
\eea
where $\alpha$ is the monopole charge density and 
$\tau$ is an arbitrary constant which represents the translational 
invariance of the magnetic field along the $z$-axis. 
Just like the monopole solution (\ref{mono}) the above
monopole string forms a classical solution of $SU(2)$ QCD,
because it satisfies the equation of motion (\ref{eom}).
But in the following we will assume $\alpha=1$ and $\tau=0$ 
for simplicity, since this will not affect the stability
analysis. Notice that here $\alpha$ has the dimension
of a mass, so that setting $\alpha=1$ amounts to fixing the scale 
in the unit of $1/\alpha$

Again we consider the eigenvalue problem
for the operator $K$
\bea
K\Psi(\rho,\phi,z) = E \Psi(\rho,\phi,z).
\label{mseveq}
\eea
With 
\bea
\Psi=\dfrac{}{}\sum_{m=-\infty}^{+\infty}F_m(\rho,z) \Phi_m(\phi),
\eea 
and repeating the steps of the previous section we
obtain the following eigenvalue equation
\bea
F_{\rho \rho} + \dfrac{1}{\rho} F_\rho
+ F_{zz} - \Big[\dfrac{(m-z)^2}{\rho^2}
\pm \dfrac{2}{\rho} - E \Big] F =0. 
\label{1aximon}
\eea
One can put $m=0$ (shifting $z$ to $z + m$) and simplify 
the equation 
\bea
F_{\rho \rho} + \dfrac{1}{\rho} F_\rho  + F_{zz} 
-\Big[\dfrac{z^2}{\rho^2} \pm \dfrac{2}{\rho} - E \Big] F =0. 
\label{eqnax2}
\eea
The quantum mechanical potential of this equation behaves like
$\pm 2/\rho$ near $\rho=0$. So we still have an undesired 
attractive potential $-2/\rho$. This implies two things. 
First, the attractive interaction of the axially symmetric monopole string
background is less severe than the attractive interaction of the 
spherically symmetric monopole background. So we can expect 
the absence of continuous negative energy spectrum 
for the axially symmetric monopole string background.
Secondly, the attractive potential $-2/\rho$ tells that 
the monopole string background must still be unstable, because 
it indicates the existence of discrete bound
states with negative energy.

To confirm this we make a qualitative estimate 
of the negative energy eigenvalues of (\ref{eqnax2}).
We look for a solution which has the form
\bea
&F(\rho,z) = \dfrac{}{} \sum_{n=0}^\infty f_n(\rho) 
Z_n(x), \nn\\
&Z_n (x) = \exp (-\dfrac{x^2}{2}) H_n(x), \nn\\
&x = \dfrac{z}{\sqrt \rho},
\label{eqnax3}
\eea
where $H_n(x)$ is the Hermite polynomial. Notice that $Z_n(x)$
forms a complete set of eigenfunctions of the harmonic oscillator,
\bea
&\Big(\dfrac{d^2}{dx^2} -x^2 \Big) Z_n(x)
=- (2n+1) Z_n(x).
\eea
Substituting (\ref{eqnax3}) into (\ref{eqnax2})
we obtain
\bea
&\dfrac{}{} \sum_{n=0}^{\infty} \Big\{\Big(\dfrac{d^2 f_n}{d\rho^2} 
+ \big(\dfrac{1}{\rho}+\dfrac{z^2}{\rho^2} \big)\dfrac{d f_n}{d\rho}
\Big) H_n \nn\\
&- \dfrac{z}{\rho \sqrt \rho} \dfrac{df_n}{d\rho}\dfrac{dH_n}{dx}  \nn\\
&+f_n \Big(\dfrac{z^2}{4 \rho^3} \dfrac{d^2 H_n}{dx^2}
-\dfrac{z}{4\rho^2 \sqrt \rho} \big(1-\dfrac{z^2}{\rho^2}\big)
\dfrac{d H_n}{dx} \Big)  \nn\\
&+ \Big(\dfrac{z^4}{4 \rho^4}- \dfrac{z^3}{4\rho^3 \sqrt \rho} 
+ \dfrac{z^2}{\rho^3} -\dfrac{2n+1\pm2}{\rho} + E \Big) f_n H_n \Big\}\nn\\
&=0.
\label{eqnlong}
\eea
Using the reccurence relations
and orthogonality properties of Hermite polynomials
one can derive the equations for $f_n(\rho)$ from (\ref{eqnlong})
\bea
&\Big(\dfrac{d^2}{d\rho^2} + \dfrac{2n+3}{2\rho} \dfrac{d}{d\rho} 
+\dfrac{4n^2-2n-1}{16 \rho^2} \nn\\
&-\dfrac{2n+1 \pm 2}{\rho} + E \Big) f_n \nn\\
&= - \dfrac{1}{64 \rho^2} f_{n-4}+\dfrac{1}{32 \rho^2} f_{n-3}
-\dfrac{1}{4\rho} \Big(\dfrac{d}{d\rho}
+ \dfrac{n-1}{4 \rho} \Big) f_{n-2} \nn \\
&+\dfrac{3}{16 \rho^2} f_{n-1}
+\dfrac{3(n+1)^2}{8 \rho^2} f_{n+1}  \nn \\
&+\dfrac{(n+1)(n+2)}{\rho} \Big(\dfrac{d}{d\rho}
-\dfrac{3n+2}{4\rho} \Big) f_{n+2}  \nn \\
&+\dfrac{(n+1)(n+2)(n+3)}{4\rho^2} \Big(f_{n+3} \nn \\
&- 3(n+4) f_{n+4} \Big), 
\eea
where $f_n=0$ for negative integers $n$. So we have infinite
number of equations for infinite number of unknown functions $f_n(\rho)$.

Notice that the left hand side of the last equation
is a second order differential equation for $f_n$ with the
quantum mechanical potential 
\bea
U = \dfrac{2n+1 \pm 2}{\rho}.
\eea
The potential becomes attractive only if $n=0$, so that  
in a first approximation we can expect that the negative energy 
eigenvalues will originate mainly from the lowest bound state 
with $n=0$ of the harmonic oscillator
part. So we can hope that by neglecting all $f_n$ with $n\neq 0$
we can still get an approximate qualitative solution
for $f_0$. In this approximation the equation reduces to 
the following simple equation
\bea
&\Big(\dfrac{d^2}{d\rho^2} + \dfrac{3}{2\rho} \dfrac{d}{d\rho}
+\dfrac{1}{\rho} -\dfrac{1}{16 \rho^2} + E \Big) f_0 =0.
\eea
The solution to this equation has a new integer
quantum number $k$, 
\bea
& f_{0,k}(\rho) =  \rho^s \exp \big(-\sqrt {|E_k|} \rho \big) 
\dfrac{}{} \sum_{l=0}^{l=k} a_l \rho^l, \nn \\
&s= \dfrac{\sqrt 2 -1}{4}, \nn \\
& E_k = -\dfrac{1}{(2k + 2s + 3/2)^2}, \nn \\
& a_{l+1} = \dfrac{\sqrt {|E_k|} (2l + 2s + 3/2)-1}{(l+1)
(l+2s + 3/2)} a_l, 
\label{eqng0}
\eea                                                                           
With this we may express the corresponding eigenfunction 
$\Psi_k$ as
\bea
& \Psi_k (\rho,\phi,z)= N_k \exp \big(-\dfrac{z^2}{2\rho} \big) 
f_{0,k}(\rho),
\label{solution}
\eea
where $N_k$ is a normalization constant. From this we find the 
lowest energy eigenvalues as follows 
\bea
&& E_0 = -0.343..., \nn \\
&& E_1 = -0.073...,\nn \\
&& E_2 = -0.031...,\nn \\
&& E_3 = -0.017..., \nn \\
&& E_4 = -0.011....
\label{spectr}
\eea                                                                           
This confirms that the axially symmetric monopole string background
is indeed unstable.

Surprisingly, we find that the approximate solution (\ref{solution})
can also be obtained as an exact solution of variational method
with the trial function $\tilde F$ of the form
\bea
&\tilde F(\rho,z) = N \rho^s \exp \big(-\beta_k \rho-\gamma 
\dfrac{z^2}{2\rho} \big) \nn\\
&\times \dfrac{}{}\sum_{l=0}^{l=k} a_l \rho^l, 
\eea
where $s,\beta_k,\gamma,a_l$ are 
treated as variational parameters.
In other words, the variational minimum of the energy functional 
with the above trial function is provided exactly 
by the solution (\ref{solution}) with (\ref{spectr}). 
                                                                                
The knowledge of the solution (\ref{solution}) allows us to develope
the perturbation theory to find a more accurate solution.
To do this we split the original equation (\ref{1aximon}) into two parts
\bea
&(H_0 + H_1)F(\rho,x) = -E F(\rho,x), \nn\\
& H_0 = \pro_{\rho\rho} + \dfrac{3}{2\rho} \pro_\rho 
+ \dfrac{1}{\rho} -  \dfrac{1}{16 \rho^2} 
+ \dfrac{1}{\rho} \pro_{xx} + \dfrac{1-x^2}{\rho}, \nn\\
& H_1 = \dfrac{x^2}{4 \rho^2} \pro_{xx} + \dfrac{x}{4 \rho^2} \pro_x 
-\dfrac{x}{\rho} \pro_{\rho x} 
-\dfrac{1}{2\rho} + \dfrac{1}{16 \rho^2}, 
\label{schr2}
\eea
and treat $H_1$ as a perturbation.
Looking for eigenfunctions of
the non-perturbed Hamiltonian $H_0$ in the form of
\bea
&F(\rho,x) = f(\rho) Z(x),
\eea 
we can separate the variables $\rho$ and $x$
\bea
&\Big(\pro_{\rho \rho} + \dfrac{3}{2\rho} \pro_\rho
-\dfrac{1}{16 \rho^2} + \dfrac{1-2n}{\rho} \Big) f(\rho) 
= - E f(\rho), \nn\\
&(\pro_{xx} - (1-x^2))Z_n(x) = - 2n Z_n(x), \nn \\
&Z_n(x) = \exp \big(-\dfrac{x^2}{2} \big) H_n(x), \nn\\
&(n=0,1,2,...).
\eea
The equation for $f(\rho)$ has solutions with a discrete negative
spectrum when $n=0$ and continious positive spectrum
when $n$ is a positive integer.
For $n=0$ the corresponding eigenfunctions $f_{0,k}(\rho)$
and discrete  energy spectrum $E_k$
are given by (\ref{eqng0}) and(\ref{spectr}).
For positive integer $n$ the physical solution is numerated
by $n$ and continious positive parameter $E$, and expressed in terms of
the confluent hypergeometric function
${\cal F}(a,b,\zeta)$ of Kummer
\bea
& f_{n,E} (\rho) = \exp \big(-i\sqrt E \rho \big) 
\rho^{\frac{\sqrt 2 -1}{4}} \nn \\
&\times {\cal F}(\dfrac{\sqrt 2+2}{4} + i \dfrac{2n+1}{2 \sqrt E}, 
\dfrac{\sqrt 2+1}{2}, 2i \sqrt E \rho). 
\eea
Notice that this solution is real and
has a correct finite limit at $E=0$ and
correct asymptotic behaviour at $\rho = \infty$
\bea
&f_{n,E} \rightarrow \rho^{-\frac{3}{4}}.
\eea
The eigenfunctions $f_{n,E}$ can be normalized
with the standard normalization prescription
\bea
\int_0^\infty f_{n,E'}^* f_{n,E}  \rho^{\frac{3}{2}} d\rho 
= 2 \pi \delta(E'-E).
\eea
The solution $f_{n,E}$ is numbered by
two quantum numbers $n$ and $E$, where $E$
takes discrete values $E=E_k$ for negative energies and continious
values for positive ones, and forms a complete set of 
eigenfunctions.

In the lowest order of perturbation theory we have, 
for the bound states (i.e., for $n=0$), 
the same energy spectrum $E_k$ as given by 
(\ref{eqng0}). The higher order corrections
to the energy eigenvalues $E_k = E_{n=0,E=E_k}$ of the bound states
are given by the perturbative calculation
\bea
& E^{(1)}_{0,E_k} = <0,E|H_1|0,E> \nn \\
&= \dfrac{}{}\int \rho^{\frac{3}{2}} f_{0,E}^{*}(\rho,x) f_{0,E}(\rho,x) 
d\rho dx d\phi, \nn\\
& E^{(2)}_{0,E_k} = \dfrac{}{}\int \dfrac {dE'}{E_k - E'} \nn\\
&\times \dfrac{}{}\sum_n <0,E_k|H^\dagger_1|n',E'><n',E'|H_1|0,E_k>.
\eea
Notice that all first order corrections $E^{(1)}_{0,E_k}$
are vanishing identically due to the fact that the solution (\ref{solution})
turns out to be an exact solution of the variational method.
Solving numerically the equation (\ref{schr2})
we obtain
\bea
&& E_0 = -0.545..., \nn \\
&& E_1 = -0.093...,\nn \\
&& E_2 = -0.036...,\nn \\
&& E_3 = -0.019..., \nn \\
&& E_4 = -0.011....
\eea
As we can see the higher order corrections change 
the ground state energy most significantly, 
but the qualitative features of the 
solution remain the same. From this we conclude that
the approximate solution (\ref{solution}) 
provides a good qualitative
estimation of the energy spectrum for us to analyse the
vacuum stability of the axially symmetric monopole string 
background.
                                                                                
The lesson that we learn from this analysis is that
the monopole string background is still unstable, but
it does improve the instability of the spherically symmetric
monopole background significantly. This is because here the
unstable attractive force becomes milder. This raises a hope 
that a pair of the axially symmetric monopole and anti-monopole
strings might form a stable magnetic background. In the following 
we prove that this is indeed the case. 

\section{A Stable Magnetic Background: Axially Symmetric 
Monopole String and Anti-monopole String Pair}
                                                                                
The main idea how to construct a stable magnetic background 
is straightforward now. Consider a pair of axially symmetric monopole
and anti-monopole strings which are orthogonal
to the $xy$-plane and separated by a distance $a$. 
Now, due to the opposite directions of the magnetic fields
of the monopole and anti-monopole strings
the total quantum mechanical potential $U(\rho)$ in the
eigenvalue equation falls down when $\rho \rightarrow \infty$
as $U(\rho) \rightarrow \,\, O(1/\rho^2)$.
By decreasing the distance $a$ we can decrease
the effective size of the quantum mechanical
potential well (as we will show below),
so that the bound state energy levels
will be pushed out from the well at some
finite critical value of $a$. This implies that the bound states will
have disappeared completely at small enough $a$.

To show this, consider an axially symmetric monopole and 
anti-monopole string pair located at $(\rho=a/2,\phi=0)$
and $(\rho=a/2,\phi=\pi)$ in cylindrical coordinates.
The magnetic field strengths  ${\vec H}_{\pm}$ for
the monopole and anti-monopole strings are given by
\bea
& \vec H_{\pm} = \pm \dfrac{\alpha}{\rho_{\pm}} \hat \rho_{\pm}, \nn \\
& \vec \rho_{\pm}= \vec \rho \pm \dfrac{\vec a}{2}, \nn \\
& \rho_{\pm}^2 =\rho^2 \pm a \rho \cos \phi +\dfrac{a^2}{4}, 
\eea
where $\vec a$ is the two-dimensional vector starting from 
the anti-monopole string to the monopole string in $xy$-plane.
Again from now on we will assume $\alpha=1$ without loss of generality.
The total magnetic field is given by
\bea
&\vec H = \vec H_+ + \vec H_-, \nn \\
&H_{\rho} = \dfrac{\rho-\dfrac{a \cos \phi}{2}}{\rho_+^2} 
-\dfrac{\rho+ \dfrac{a \cos \phi}{2}}{\rho_-^2}, \nn \\
&H_{\phi} = \dfrac{a \rho(\rho^2 + \dfrac{a^2}{4}) \sin \phi}
{\rho_+^2 \rho_-^2},
~~~~~H_{z} =0 , \nn \\
&H = \sqrt {H_\rho^2 + \dfrac{H_\phi^2}{\rho^2}} 
= \dfrac{a}{\rho_+ \rho_-}.
\eea                                                                           
The corresponding vector potential is given by
\bea
&\tC_\mu = z H_{\phi} \pro_\mu \rho
-\rho z H_{\rho} \pro_\mu \phi.
\eea
The eigenvalue equation for the operator $K$ has the form
\bea
&\Big\{-\dfrac{1}{\rho} \pro_\rho (\rho \pro_\rho )
- \dfrac{1}{\rho^2} \pro_\phi^2 - \pro_z^2
- 2i\dfrac{z}{\rho} \Big(H_{\phi} \pro_\rho 
-H_{\rho} \pro_\phi \Big) \nn\\
&+ z^2 H^2 \pm 2H \Big\} \Psi(\rho,\phi,z) = E \Psi(\rho,\phi,z). 
\label{orig}
\eea
The equation can be interpreated as a Schroedinger equation
for a massless gluon in the magnetic field of monopole and
anti-monopole string pair.
                                                                                
Let us analyse the equation qualitatively first.
We will concentrate on the potentially dangerous 
$-2H$ potential in (\ref{orig}).
The singularities of the term $z^2 H^2$
determine the essential singularitites of the differential equation.
One can try to extract the leading factor of the solution
and look for a finite solution for the
ground state in the form
\bea
\Psi(\rho,\phi,z) = (\pi \rho_+ \rho_-)^{-\frac{1}{4}}
\exp\big(-\dfrac{z^2}{2 \rho_+ \rho_-}\big)
F(\rho,\phi),   
\label{solform}
\eea
where $F(\rho,\phi)$ is normalized
\bea
\int |F(\rho,\phi)|^2 \rho d\rho   d\phi = 1.
\eea
The solution describes a wave function localized mainly near
the string pair. The wave function vanishes exactly on 
the axes of the strings. This implies that the ground state 
has a non-zero orbital angular momentum
which provides a centrifugal potential as we will see
later.
                                                                               
The lowest negative eigenvalue of this equation can be
obtained by variational method by minimizing the corresponding
energy functional
\bea
&E=\dfrac{}{} \int \Psi^* \Big [-\dfrac{1}{\rho} \pro_\rho (\rho \pro_\rho )
-\dfrac{1}{\rho^2} \pro_\phi^2 - \pro_z^2 
- 2i\dfrac{z}{\rho} (H_\phi \pro_\rho \nn\\
&- H_\rho \pro_\phi) 
+ z^2 H^2 \pm 2H \Big ] \Psi \rho d\rho dz d\phi.
\eea
Now, with
\bea
F(\rho,\phi)= \dfrac{}{} \sum_{-\infty}^{+\infty}
f_m (\rho) \Phi_m,
\eea
one may suppose that the main contribution
to the ground state energy comes from the first term of Fourier
expansion with $m=0$. With this we can perform the integration 
over $z$-coordinate, and simplify the above expression to 
\bea
&E =\dfrac{}{} \int f(\rho) \Big(-f''(\rho)
-\dfrac{1}{\rho} f'(\rho) \nn \\
& + U(\rho,\phi) f(\rho)\Big) \rho d\rho d\phi, \nn\\
& U(\rho,\phi) = \dfrac{\rho^2 - 2 a \rho_+ \rho_-}{2 \rho_+^2\rho_-^2},
\eea
where $f(\rho)=f_0(\rho)$ and $U(\rho,\phi)$ is an effective 
potential. To understand the nature of the potential
we make the following rescaling
\bea
&\rho \rightarrow a \rho,  
~~~~~f \rightarrow f/a,  \nn\\
&E \rightarrow E/a^2,
\eea
and find that under this rescaling the potential near the origin
can be approximated to
\bea
U(\rho,\phi) \rightarrow -4 a
+ (8 -16a \cos 2\phi) \rho^2.
\label{near0}
\eea
So after the rescaling the potential reduces to a two dimensional 
harmonic oscillator potential whose depth 
decreases as $a$ goes to zero. 
This implies that the negative energy eigenvalues 
will disappear for $a$ less than a 
finite critical value.

To complete our analysis we perform
the integration over the angle variable $\phi$
in the energy functional, and with the change of variable
\bea
f(\rho) = \chi (\rho) /\sqrt{\rho},
\eea
we obtain the following equation 
which minimizes the energy, 
\bea
&\Big[-\dfrac{d^2}{d\rho^2}+ V(\rho) \Big] \chi(\rho) = E \chi(\rho), \nn\\
& V(\rho) = -\dfrac{1}{4\rho^2}+ \dfrac{8\rho^2}
{\sqrt{(a^4-16 \rho^4)^2}} \nn \\
&-\dfrac{8a}{\pi \sqrt {(a^2-4\rho^2)^2 }} 
K\big(-\frac{16 a^2 \rho^2}{(a^2 - 4 \rho^2)^2}\big),
\label{eq10}
\eea
where $K(x)$ is the complete elliptic integral of the first kind.
With this we have the asymptotic behavior of
the potential $V(\rho)$ at space infinity
\bea
V(\rho) \simeq (\dfrac{1}{4}-a) \dfrac{1}{\rho^2}.
\eea
This tells that the potential becomes positive when the distance
$a$ becomes less than the critical value $a_0$ (in the unit $1/\alpha$)
\bea
a<a_{0}=\dfrac{1}{4}.
\eea
A careful numerical analysis of (\ref{orig}) gives us 
the following critical value 
\bea
a_{0}\simeq 0.24,
\eea
which is close to the approximate value $1/4$.
This confirms that qualitatively the approximate solution
describes the correct physical picture. In particular,
this tells that a pair of monopole and antimonopole strings
become a stable magnetic background if the distance between
two strings is small enough. 
                                                                                
\section{Instability of Magnetic Vortex-Antivortex Pair}
                                                                                
Recently an alternative mechanism of confinement has been proposed
which advocates the condensation of magnetic vortices \cite{diak}. 
However, it has been known 
that the magnetic vortex configuration
is unstable \cite{bordag}. So it would be interesting to 
study the stability of the vortex-antivortex pair.
In this section we study the stability of 
the vortex and anti-vortex pair and show that the pair is unstable.
                                                                                
Let us start with a single vortex configuration
given by 
\bea
&\vec H= \dfrac{1}{\rho} \hat z,
~~~~~\tC_\mu =\rho \pro_\mu \phi.
\eea
Notice that, unlike the monopole string (\ref{mstring}), 
the vortex configuration is not a classical solution of the system. 
But since this type of configuration multiplied by 
an appropriate profile function has been studied by many 
authors \cite{diak}, we will
consider the vortex configuration in the following.

The corresponding eigenvalue equation of the operator
$K$ is given by
\bea
&\Big[-\pro_\rho^2-\dfrac{1}{\rho} \pro_\rho - \dfrac{1}{\rho^2} \pro_\phi^2
-\pro_z^2 -\dfrac{2i}{\rho} \pro_\phi 
\pm 2 H \Big] \Psi(\rho,\phi,z) \nn\\
&= E \Psi(\rho,\phi,z).
\label{schr3}
\eea
The equation becomes separable in all three variables.
With 
\bea
&\Psi= \dfrac{}{} \sum_{-\infty}^{+\infty}
f_m(\rho) g(z) \Phi_m(\phi),~~~~~g(z)=1,
\eea
one obtains the following ordinary differential
equation for $f(\rho)$ from (\ref{schr3}),
\bea
&\Big(-\pro_\rho^2-\dfrac{1}{\rho} \pro_\rho
+ (1+\dfrac{m}{\rho})^2 \pm \dfrac{2}{\rho} - E\Big) f(\rho) \nn\\
&=0.
\eea
The bound states are possible for the potential $-2/\rho$
with non-positive integer $m$, 
in which case the corresponding solution
is given by
\bea
& f_{n,m}(\rho) = \rho^{|m|} e^{-\sqrt{1-E_{n,m}}} u_{n,m}(\rho), \nn \\
& u_{n,m} (\rho) = \dfrac{}{}\sum_{k=0}^n a^{n,m}_k \rho^k, \nn \\
& a^{n,m}_{k+1} = \nn\\
&\dfrac{\sqrt{1-E_{n,m}} (2 k +2 |m|+1) -2+2m}
{(k+1)(k+2 |m| +1)} a^{n,m}_k, \nn \\
& E_{n,m} = 1-\dfrac{4 (1-m)^2}{(2n+2 |m| +1)^2}, \nn \\
& n=0,1,2,...; \,\, m=0,-1,-2,...
\eea
Clearly the ground state has a negative energy $E_{0,0}$,
which tells that the vortex configuration is unstable.

There are two principal differences between 
the vortex configuration and
the axially symmetric monopole string. First, the monopole string
is a classical solution of $SU(2)$ QCD, but the vortex configuration 
is not. Secondly, the ground state 
eigenfunction $f_{0,0}(\rho)$ of the vortex configuration 
corresponds to an $S$-state, 
which implies the absence of the centrifugal
potential. But the ground state of the monopole string
has a non-trivial centrifugal
potential. As we will see soon this will play the important role
in the existence of the negative energy eigenstates in the case
of vortex-antivortex background.
                                                                                
The vortex-antivortex background is described in
a similar manner as the monopole-antimonopole string 
background in the last section. The potential is given by
\bea
&\tC_\mu = \dfrac{a}{2} \sin \phi 
(\dfrac{1}{\rho_+}+\dfrac{1}{\rho_-}) \pro_\mu \rho \nn \\
&+ \Big[\dfrac{\rho}{\rho_+} (\rho+\dfrac{a}{2} \cos \phi) 
-\dfrac{\rho}{\rho_-}(\rho-\dfrac{a}{2} \cos \phi) \Big] \pro_\mu \phi, \nn\\
& \vec H = \Big(\dfrac{1}{\rho_+} - \dfrac{1}{\rho_-} \Big) \hat z,
\eea
where $a$ is the distance between the axes of the vortex and anti-vortex.
The eigenvalue equation corresponding to the
operator $K$ is given by
\bea
&\Big[-\pro_\rho^2-\dfrac{1}{\rho} \pro_\rho 
-\dfrac{1}{\rho^2} \pro^2_\phi-\pro_z^2
- 2i (\tC_{\rho} \pro_\rho+\dfrac{1}{\rho^2} \tC_{\phi} \pro_\phi)\nn \\
&+\tC_\mu^2 \pm 2H \Big] F(\rho,\phi)=E F(\rho,\phi),
\eea
which is partially factorizable in $z$-coordinate.
The numerical analysis of the equation shows that there is no
critical value for the parameter $a$, so that
the negative energy eigenvalues exist for any small $a$. Qualitatively
one can see this from the effective the potential
$\tC_\mu^2-2H$. After averaging over the angle variable the potential
becomes
\bea
&V(\rho)= 2 \dfrac{}{} \int_0^{2 \pi} 
(1- |\dfrac{1}{\rho_+}-\dfrac{1}{\rho_-}|
-\dfrac{\rho^2-\dfrac{a^2}{4}}{\rho_+ \rho_-}) d\phi  \nn\\
&=4 \pi+\dfrac{16}{\rho+\dfrac{a}{2}} F(\dfrac{\pi}{4},x_{-}) 
-\dfrac{16}{|\rho-\dfrac{a}{2}|} F(\dfrac{\pi}{4},-x_{-}) \nn \\
&-\dfrac{8}{\rho+\dfrac{a}{2}} K(x_{+}) 
+\dfrac{8}{|\rho-\dfrac{a}{2}|} K(-x_{-})  \nn \\
&-\dfrac{8 (\rho-\dfrac{a}{2})}{|\rho-\dfrac{a}{2}|}
K(-\dfrac{x_{+}x_{-}}{4}), \nn\\
&x_{\pm}= \dfrac{2a \rho}{(\rho \pm \dfrac{a}{2})^2}
\eea
where $F(w,x), K(x)$ are the elliptic and the complete elliptic
integral of the first kind.
So near the origin and infinity we have
\bea
& V(\rho)\simeq \left\{{\dfrac{}{}8\pi-\dfrac{64}{a^2} \rho 
- \dfrac{16 \pi}{a^2} \rho^2~~~(\rho \simeq 0), 
\atop ~-\big(\dfrac{}{}4-\dfrac{\pi}{2}a \big) \dfrac{2a}{\rho^2} 
~~~~~~~(\rho \simeq \infty).}\right.
\eea
This shows that the effective potential has no centrifugal 
potential which could prevent the appearance of bound states
for small $a$.
This is the origin of existence of the negative energy eigenvalues
and instability of the vortex-antivortex background.
Whether the instability problem can
be overcome with a more complicate configuration
of the vortex-antivortex is an open and interesting question.
                                                                                
\section{Conclusions}
                                                                                
In this paper we have shown that the
axially symmetric monopole-antimonopole string background
is stable under the quantum fluctuation, if the distance 
between two string becomes less than the critical value 
$a_0 \simeq 1/4$. As far as we understand it this is the first
explicit example of a stable magnetic background in $SU(2)$ QCD.
The existence of the stable magnetic background strongly implies
that ``a spagetti of gauge invariant monopole-antimonopole string
pairs" could generate a stable vacuum condensation in QCD. 
This would allow a magnetic confinement of color in QCD. 

Another important result of this paper is that a pair of 
magnetic vortex and anti-vortex strings is unstable.
In the magnetic confinement mechanism there have been two competing
ideas, the monopole-antimonopole condensation and 
the magnetic vortex-antivortex condensation, and recently
the magnetic vortex-antivortex condensation has been advocated 
by many authors as a possible confinement mechanism 
in QCD \cite{diak}. Our result suggests that 
the magnetic vortex-antivortex condensation is not likely
to generate a stable magnetic vacuum.

The search for the existence of a stable
magnetic condensation in QCD has been painful.
Savvidy first calculated the one-loop effective action
of $SU(2)$ QCD with a constant magnetic background.
But the Savvidy background was unstable, 
because it was not gauge invariant. In fact any classical background
which is not gauge invariant can not be physical,
and thus can not be a stable vacuum.
Because of this Nielsen and Olesen have proposed the gauge invariant
``Copenhagen vacuum", and conjectured that such a gauge invariant 
vacuum must be stable under the quantum fluctuation \cite{niel}.
Although conceptually very attractive the ``Copenhagen vacuum", 
however, was not so useful in practical purposes.

But recently it has been shown that if we impose the gauge 
invariance to the SNO vacuum, the imaginary part of the effective action 
disappears \cite{qcd8}. This tells that if we imposes
the gauge invariance properly, QCD can generates a stable 
magnetic condensation. The result in this paper strongly endorses
this. Although the axially symmetric monopole-antimonopole string
background can not be identified as a vacuum because it is not
translationally invariant, it clearly indicates the existence of 
a stable magnetic vacuum. Furthermore, it indicates that a stable 
magnetic vacuum must be, not just a condensation of monopoles
but, a condensation of a gauge invariant combination of
monopoles and anti-monopoles. In fact our result strongly suggests
that a ``spagetti of monopole-antimonopole string pairs",
or more precisely an infinite set of monopole-antimonopole 
string pairs which forms a square crystal in $xy$-plane,
can be a stable QCD vacuum if the size of the crystal 
is small enough.

Finally we wish to point out a possible connection between
our monopole-antimonopole string background and the 
``Copenhagen vacuum". The ``Copenhagen vacuum" can be viewed as 
a collection of domains of constant magnetic fields 
which have random directions.
Our result suggests that an infinite set 
of monopole-antimonopole string pairs which forms 
a square crystal in $xy$-plane can be viewed as
a stable vacuum, and thus a candidate of ``Copenhagen vacuum". 
In this sense our result 
strongly supports the idea of  ``Copenhagen vacuum". 
On the other hand, our result also imposes a strong
constraint on the ``Copenhagen vacuum". 
In a simplified picture the ``Copenhagen vacuum" was proposed 
as a ``spagetti of colored magnetic tubes" \cite{niel}. 
But our result shows that the magnetic vortex-antivortex pair is unstable. 
This implies that ``spagetti of colored magnetic tubes" may not 
be viewed as a ``Copenhagen vacuum". Rather, a ``spagetti of 
monopole-antimonopole string pairs" forms a stable magnetic vacuum,
and thus can be viewed as a ``Copenhagen vacuum".
This is consistent with the picture of ``gauge invariant"
monopole condensation.
 
{\bf Acknowledgements}
                                                                                
~~~One of the authors (YMC) thanks G. Sterman for the kind hispitality
during his visit to C.N. Yang Institute of Theoretical Physics.
The work is supported in part by the ABRL Program of
Korea Science and Engineering Foundation (R14-2003-012-01002-0) 
and by the BK21 Project of the Ministry of Education.


\begin{thebibliography}{99}
                                                                                
\bibitem{nambu}Y. Nambu, Phys. Rev. {\bf D10}, 4262 (1974);
S. Mandelstam, Phys. Rep. {\bf 23C}, 245 (1976);
A. Polyakov, Nucl. Phys. {\bf B120}, 429 (1977);
G. 't Hooft, Nucl. Phys. {\bf B190}, 455 (1981).
\bibitem{cho1}Y. M. Cho, Phys. Rev. {\bf D21}, 1080 (1980);
J. Korean Phys. Soc. {\bf17}, 266 (1984).
\bibitem{cho2}Y. M. Cho, Phys. Rev. Lett. {\bf 46}, 302 (1981);
Phys. Rev. {\bf D23}, 2415 (1981); Z. Ezawa and A. Iwazaki,
Phys. Rev. {\bf D25}, 2681 (1982).
\bibitem{savv} G. K. Savvidy, Phys. Lett. {\bf B71}, 133 (1977);
S. G. Matinyan and G. K. Savvidy, Nucl. Phys. {\bf B134}, 539 (1978);
N. K. Nielsen and P. Olesen, Nucl. Phys. {\bf B144}, 376 (1978);
\bibitem{ditt} A. Yildiz and P. Cox, Phys. Rev. {\bf D21}, 1095 (1980);
M. Claudson, A. Yilditz, and P. Cox, Phys. Rev. {\bf D22}, 2022 (1980);
W. Dittrich and M. Reuter, Phys. Lett. {\bf B128}, 321, (1983);
C. Flory, Phys. Rev. {\bf D28}, 1425 (1983);
S. K. Blau, M. Visser, and A. Wipf, Int. J. Mod. Phys.
{\bf A6}, 5409 (1991); M. Reuter, M. G. Schmidt, and C. Schubert,
Ann. Phys. {\bf 259}, 313 (1997).
\bibitem{niel} H. B. Nielsen and M. Ninomiya,
Nucl. Phys. {\bf B156}, 1 (1979);
N. K. Nielsen and P. Olesen, Nucl. Phys. {\bf B160},
380 (1979); C. Rajiadakos, Phys. Lett. {\bf B100}, 471 (1981).
\bibitem{sch} V. Schanbacher, Phys. Rev. {\bf D26}, 489 (1982).
\bibitem{cho3} Y. M. Cho, H. W. Lee, and D. G. Pak,
Phys. Lett. {\bf B 525}, 347 (2002); Y. M. Cho and D. G. Pak,
Phys. Rev. {\bf D65}, 074027 (2002).
\bibitem{cho99} Y. M. Cho and D. G. Pak,
in {\it Proceedings of TMU-YALE
Symposium on Dynamics of Gauge Fields}, edited by T. Appelquist and H.
Minakata (Universal Academy Press) Tokyo, 1999;
Y. M. Cho and D. G. Pak, J. Korean Phys. Soc. {\bf 38}, 151 (2001).
\bibitem{cho4} Y. M. Cho, D. G. Pak, and M. Walker,
JHEP {\bf 05}, 073 (2004);Y. M. Cho and M. L. Walker,
Mod. Phys. Lett. {\bf A19}, 2707 (2004).
\bibitem{qcd8} Y. M. Cho, hep-th/0301013.
\bibitem{wu} T. T. Wu and C. N. Yang, Phys. Rev. {\bf D12}, 3845 (1975);
T. T. Wu and C. N. Yang, Nucl. Phys. {\bf B 107}, 365 (1976);
Phys. Rev. {\bf D 16}, 1018 (1977).
\bibitem{cho80} Y. M. Cho, Phys. Rev. Lett. {\bf 44}, 1115 (1980);
Phys. Lett. {\bf B115}, 125 (1982).
\bibitem{brandt} R. A. Brandt and F. Neri, Nucl. Phys. {\bf B 161},
253 (1979).
\bibitem{fadd} L. Faddeev and A. Niemi, Phys. Rev. Lett.
{\bf 82}, 1624 (1999); Phys. Lett. {\bf B449}, 214 (1999).
\bibitem{lang}E. Langman and A. Niemi, Phys. Lett. {\bf B463}, 252 (1999);
S. Shabanov, Phys. Lett. {\bf B458}, 322 (1999); {\bf B463}, 263 (1999);
H. Gies, Phys. Rev. {\bf D63}, 125023 (2001).
\bibitem{zucc} R. Zucchini, Int. J. Geom. Meth. Mod. Phys. {\bf 1}, 813
(2004).
\bibitem{kondo} K. Kondo, Phys. Lett. {\bf B600}, 287 (2004);
hep-th/0410024; K. Kondo, T. Murakami, and T. Shinohara, hep-th/0504107.
\bibitem{cho00} Y. M. Cho, Phys. Rev. {\bf D62}, 074009 (2000).
\bibitem{cho01} W. S. Bae, Y. M. Cho, and S. W. Kimm, 
Phys. Rev. {\bf D65}, 025005 (2001).
\bibitem{dewitt}B. de Witt, Phys. Rev. {\bf 162}, 1195 (1967);
1239 (1967).
\bibitem{pesk} See for example,
C. Itzikson and J. Zuber, {\it Quantum Field Theory} (McGraw-Hill) 1985;
M. Peskin and D. Schroeder,
{\it An Introduction to Quantum Field Theory} (Addison-Wesley) 1995;
S. Weinberg, {\it Quantum Theory of Fields} (Cambridge Univ. Press) 1996.
\bibitem{schw} J. Schwinger, Phys. Rev. {\bf 82} , 664 (1951).
\bibitem{table} See, for example, I. Gradshteyn and I. Ryzhik,
{\it Table of Integrals, Series, and Products}, edited by A. Jeffery
(Academic Press) 1994; M. Abramowitz and I. Stegun,
{\it Handbook of Mathematical Functions}, (Dover) 1970.
\bibitem{landau} L. D. Landau and E. M. Lifshitz,
{\it Course of Theoretical Physics: Vol. 3, Quantum Mechanics} Pergamon Press,
1977.
\bibitem{diak} D. Diakonov and M. Maul, Phys. Rev. {\bf D66} (2002) 096004;
J. D. Lange, M. Engelhardt and H. Reinhardt, Phys. Rev. {\bf D68}, 025001
 (2003).
\bibitem{bordag} M. Bordag, Phys. Rev. {\bf D67}, 065001 (2003).
                                                                                
\end{thebibliography}
\end{document}